\documentclass[pra,aps,twocolumn,showpacs]{revtex4}
\usepackage{graphicx,psfrag}

\begin{document}
\title{Stripe Formation in  
Bose-Einstein Condensates with Large Numbers of Vortices}
\author{Erich J. Mueller and Tin-Lun Ho}
\affiliation{Department of Physics,  The Ohio State University,
Columbus, Ohio
43210}

\begin{abstract}
We analyze recent experiments at JILA which observe periodic formation of
stripes in a  rotating Bose condensate [Engels et al. 
Phys. Rev. Lett. {\bf 89}, 100403 (2002)], and explain this behavior
as a consequence of kinetic energy driven evolution,
where the vortices follows a quadrupolar flow. 
We demonstrate that
the structure of vortex
lattices in the lowest Landau level is a useful tool for understanding such complex phenomena,
even when the system is not in the quantum Hall regime. 
\end{abstract} 
\maketitle

In the last year and a half there have been remarkable developments 
in producing Bose-Einstein condensates (BEC) with large numbers of
vortices\cite{ParisMITJILA}. These experiments  provide a unique
setting for studying the dynamics of vortex
lattices.   Such experiments would be nearly impossible in a 
conventional ultra-low temperature laboratory. 
It is conceivable that even higher angular momentum states can  be
achieved, eventually producing quantum Hall physics. 

The novelty of the dynamics at high angular momentum is demonstrated in a
recent experiment at JILA\cite{JILAstripe}.  In this work, after a vortex
lattice (with rotation frequency $\Omega$) is created, the stirring force is
turned off and a small quadrupolar deformation is suddenly introduced in the
trap. The  vortex lattice is then rotating against a $static$ anisotropic
potential.  
Even for small distortions, the atom cloud undergoes significant ellipsoid
deformation. Surprisingly,  the vortex lattice transforms periodically
into a set of parallel stripes, as if the vortices ``melt" into sheets and
reappear in a coherent fashion. The stripes appear six times over the rotating
period $T=2\pi/\Omega$, whenever the basis vector of the vortex lattice
coincide with the minor axis of the cloud.

There is little question that  this intriguing behavior is a general feature of
rotating Bose gases with large angular momentum ($L$), and hence a consequence of
the unique physics in this limit. From the view point of energetics, the most
significant  
change as $L$ increases is the rising importance of kinetic energy, which
ultimately forces the system into the lowest Landau level. We quantify this
consideration by showing that 
at sufficiently high $L$, the interactions only perturbatively enter the dynamics, and
the time evolution of the vortex lattice is driven by kinetic energy.
We calculate analytically the time evolution of the condensate,
choosing
a  variational form
of the initial 
vortex lattice based on its equilibrium structure at even higher angular
momentum (i.e. the mean field quantum Hall regime)\cite{Ho}.  
Our calculation explains the origin of all the features observed in
ref.\cite{JILAstripe}. The agreements between our results and experimental
observations shows that even if the system is not in the quantum Hall regime,
the physics of lowest Landau level, which plays an important role in our calculation,
is useful for understanding
complex phenomena and provides powerful approximation schemes.

{\bf I. Kinetic evolution: Justification and evolution of density profile:}  We
are interested the solution of the Gross-Pitaevskii (GP) equation 
\begin{equation}
i\hbar \partial_{t} \Psi({\bf x},t) = ( T + V({\bf x}) + g|\Psi({\bf
x},t)|^2)
 \Psi({\bf x},t)
\label{GP} \end{equation}
with the boundary condition $\Psi({\bf x},t=0)=\Phi({\bf r})$, where 
$T = - (\hbar^2/2M){\bf \nabla}^2$, 
$\Phi({\bf r})$ is the equilibrium vortex state of a cylindrically potential,  
 $\Psi({\bf r},t)$ is the condensate wavefunction after a quadrupolar
distortion is added to the potential at $t=0$, and $V({\bf x})$ is the deformed 
trap at time $t>0$. 
We shall treat the problem as two dimensional, since the effect of the third
dimension can be obtained by Thomas-Fermi approximation as in ref.\cite{Ho}. 
In this case, we have $V =\frac{1}{2}M(\omega_{x}^2 x^2 + \omega_{y}^2 y^2)$,  
$\omega_{x} = \omega(1+ \epsilon)$,  $\omega_{y} = \omega(1- \epsilon)$, where 
$\epsilon$ is the quadrupolar distortion.  

Since $\Phi({\bf r})$ is an equilibrium rotating state, it minimizes the GP
functional ${\cal F} = \langle h_0+U -\Omega L_{z}-  \mu N\rangle$, 
where $h_0= T+V$,
$\langle U\rangle=(g/2)\int\!\! |\Phi|^4$,
$\mu$ is the chemical potential and $L_{z} = -i \hbar \hat{\bf z} \cdot {\bf x}
\times \nabla$. Writing $\Phi({\bf r})= e^{iS({\bf r})}\sqrt{\rho({\bf r})}$,
where $S$ is the phase function and 
$\rho$ is the density, we have 
${\cal F} =    \int \{   \frac{\hbar^2}{2M}[\nabla\sqrt{\rho}]^2 
+ \frac{M}{2}({\bf v}_{s} - \Omega\hat{\bf z}\times {\bf r})^2\rho  +
[\frac{1}{2} M(\omega^2 - \Omega^2) r^2  -\mu]\rho
+ \frac{1}{2}g\rho^2  \} $
where ${\bf v}_{s}= (\hbar/M)\nabla S$ is the superfluid velocity. For
equilibrium states with large number of vortices, the system has a spatial
average $\overline{ {\bf v}}_{s} \approx \Omega \hat{\bf z} \times {\bf r}$,
and a density is  
\begin{equation}
\rho_{o} \approx (\mu  - \frac{1}{2}M(\omega^2 - \Omega^2)r^2 )/g.
\label{rottf} \end{equation}
It is important to note that states with large number of vortices can only be
achieved when $\Omega\rightarrow \omega$, which means that 
 $\frac{1}{2}M(\omega^2 - \Omega^2)r^2 <<  V({\bf r})= 
 \frac{1}{2}M\omega^2 r^2 $. As a result, 
the $g|\Psi|^2$ term in Eq.~(\ref{GP}) can be replaced by a constant 
($g|\Psi|^2\approx\mu$) at 
early times where $\Psi \approx \Phi$. 
We show below, however, that this replacement remains valid over the observed 
period of stripe formation, and the dynamics are governed solely by the kinetic energy,
leading us to describe the dynamics as
{\em kinetic evolution}.

Under kinetic evolution, the shape of the condensate is described by
the matrix ${\cal D}_{ij} = \langle x_{i}(t)x_{j}(t)  \rangle /
[\langle r^2 \rangle/2]$, 
where  $(x_{1}, x_{2})\equiv (x,y)$, 
$x_{i}(t) = e^{ih_0t} x_{i}e^{-ih_0t}$,   and $\langle .. \rangle$ means average
with respect to the initial state $\Phi$.  
Since $x(t) =  c_{x} x + s_{x}
p_{x}/(M\omega_{x})$, 
$y(t) = c_{y} y + s_{y} p_{y}/(M\omega_{y})$, and ${\bf p}\Psi \approx M\Omega
\hat{\bf z} \times {\bf r} \Psi$, where $c_{x}={\rm cos}\omega_{x}t$, 
$s_{x}={\rm sin}\omega_{x}t$,  we have
$\langle x_{i}x_{j} \rangle =\frac{1}{2} \langle r^2 \rangle \delta_{ij}$, 
$\langle p_{i}p_{j} \rangle =\frac{1}{2}M^2\Omega^2 \langle r^2 \rangle
\delta_{ij} \sim \frac{1}{2}M^2\omega^2 \langle r^2 \rangle \delta_{ij}$,  
$\langle x p_{y} \rangle\approx M\Omega \langle x^2 \rangle$,  and hence
\begin{equation}
{\cal D}=   \left( 
\begin{array}[c]{cc}
c^{2}_{x}  +  t_{x}^2  &
c_{x}t_{y}  -  c_{y} t_{x}  \\
c_{x}t_{y}  -  c_{y} t_{x} &
c^{2}_{y}  +  t_{y}^2  
 \end{array} \right) 
\equiv   \left( 
\begin{array}[c]{cc} \alpha & \gamma \\
\gamma &  \beta 
\end{array} \right).
\label{D} \end{equation}
where $t_{i}= s_{i}\omega/\omega_{i}$, $i=x,y$.  (Eq. (\ref{D}) can also be
derived 
from the trial wavefunction we use later).   Noting that initial density in
Eq.~(\ref{rottf}) is of the form $\rho_{o} = \overline{\rho}
 ( 1 - r^2 /\langle r^2 \rangle)$, where $g \overline{\rho} = \frac{1}{2} M
(\omega^2 - \Omega^2)\langle r^2 \rangle$, 
Eq.~(\ref{D}) implies that with similar approximation, the deformed density is
$\rho({\bf r}) =   \frac{\overline{\rho}  }{ \sqrt{ {\rm Det}{\cal D}  }  }
\label{density} \left( 1 -  {\bf r}\cdot {\cal D}^{-1} \cdot {\bf r} /
{\langle r^2 \rangle} \right)$, 
where the inverse determinant  comes from number conservation.  The deformed
density can be rewritten as 
\begin{equation}
\rho({\bf r}) = \frac{\overline{\rho}}{\sqrt{\lambda_{+}\lambda_{-}} }
-\frac{M (\omega^2 - \Omega^2) }{2g\sqrt{\lambda_{+}\lambda_{-}}} \left( 
\frac{ r_{+}^2}{\lambda_{+}}  + 
\frac{ r_{-}^2}{\lambda_{-} } \right)   
\label{rho} \end{equation}
where $\lambda_{\pm}$ and $\hat{\bf n}_{\pm}$ are the eigenvalues and 
eigenvectors of ${\cal D}$, ($\lambda_{+}>\lambda_{-}$, 
$\hat{\bf n}_{+} \cdot \hat{\bf n}_{-}=0$),  and $r_{\pm} = {\bf r}\cdot
\hat{\bf n}_{\pm}$. 
To be able to replace $g\rho$ in Eq.~(\ref{GP}) by a constant,  one needs 
$g\rho <<\frac{1}{2}M\omega^2 (r^{2}_{+} + r^{2}_{-})$, or
\begin{equation}
1 -  \Omega/\omega << \lambda_{-} \sqrt{\lambda_{+}\lambda_{-}}. 
\label{condition} \end{equation}
In Fig. 1a, we plot the $\lambda$-product in Eq.~(\ref{condition}) and
the 
aspect ratio  $\sqrt{\lambda_{-}/\lambda_{+}}$ of the ellipsoidal density
profile
 Eq.~(\ref{rho})
as a function of $\omega t$ for $\epsilon = 0.036$. It is clear that
Eq.~(\ref{condition}) remains valid up to 2.5 to 3 times of the original period
$2\pi/\omega$ for rotating states with $1 -  \Omega/\omega \sim 10^{-2}$.  One
can see from fig.1a that 
even for $\epsilon = 0.036$, the aspect ratio reaches $50\%$ when $\omega t
\approx 
3\pi$\cite{comment1}.  In fig.1b, we have plotted the orientation of the minor
axis $\hat{\bf n}_{-}\equiv({\rm cos}\chi, {\rm sin}\chi)$, where $\chi$ can be
found from Eq.~(\ref{D}) to satisfy 
${\rm tan}2\chi = \gamma/(\alpha- \beta)$.  We see that $\hat{\bf n}_{-}$ 
starts at $\hat{\bf x}$,  and then rotate by $45^{o}$ in half a period and
stays around that value for some time. Eventually it will rotate back to $\chi
\approx 0$, but only at times much longer than that for which kinetic evolution is valid. 

For the rest of the paper we shall stay within the time interval where kinetic
evolution is valid. Eq.~(\ref{GP}) then gives \begin{equation}
\Psi({\bf x},t) = \int G({\bf x}, {\bf x'}; t) \Phi({\bf x'}) \,\,{\rm d}{\bf x'}, 
\label{int} \end{equation}
where $G({\bf x}, {\bf x'}; t)= \langle {\bf x}|e^{-i h_0 t/\hbar} |{\bf
x'}\rangle$, 
$h_0 = {\bf p}^2/2M + M(\omega^2_{x} x^2 + \omega^2_{y} y^2)/2$, 
 which can be shown to be 
\begin{eqnarray}
&G({\bf r}, {\bf r'}; t) 
= {\cal A}e^{(i/2)({\bf r}\cdot {\cal Q} \cdot {\bf r} + 
{\bf r'}\cdot {\cal Q} \cdot {\bf r'}
-2 {\bf r}\cdot  {\cal P}\cdot {\bf r'})},&
\label{green} \\
&
{\cal Q} = \frac{1}{a^2} \left( \begin{array}[c]{cc}  c_{x}/t_{x} & 0 \\
0 &  c_{y}/t_{y} \end{array}\right), 
 \,\,\,\,\,\,\,\,
{\cal P} = \frac{1}{a^2}  \left( \begin{array}[c]{cc} 1/t_{x} & 0 \\
0 &  1/t_{y} \end{array}\right)&
\label{Q} \end{eqnarray}
and ${\cal A} = (2\pi a^2)^{-1} \sqrt{ \frac{ c_{x} c_{y}  }{ - t_{x} t_{y} }  
} $, 
where $a^2= \hbar/(M\omega)$. 

{\bf II. The evolving condensate:}  For later discussions, we represent the
basis vector ${\bf b}_{1}, {\bf b}_{2}$ of a vortex lattice as complex numbers
$b_{i}= {\bf b}_{i}\cdot(\hat{\bf x} + i\hat{\bf y})$, $i=1,2$,  and define
$\tau \equiv b_{2}/b_{1}  = u+ iv$, where   $u$ and $v$ are the real and
imaginary part of $\tau$.  The entire lattice will be denoted as $\{ b \}$, 
where  $b=\ell_{1}b_{1} + \ell_{2}b_{2}$, and  $\ell_{1}, \ell_{2}$ are
integers;  and the size of unit cell is $|b_{1}|^2 |v|$.  The choice of basis is
not unique. A natural choice is the ``fundamental set"  ($\overline{b}_{1},
\overline{b}_{2}$) containing the two 
shortest non-parallel vectors in the lattice.  
As well known from Conformal Analysis,  this set is confined in the 
``fundamental" region $(|\tau|>1, |u|< 1/2, v>0)$, and that for 
any basis ($b_{1}, b_{2}$)  one can find an {\em image}  ($\overline{b}_{1},
\overline{b}_{2}$) in the fundamental region by basis transformation.  Because
the way the basis vectors are chosen, the fundamental region is best for
describing the stretching of lattices. 
 To distinguish the initial hexagonal lattice from others at later time, we
label its quantities with a superscript ``$^{o}$". Thus, $\tau^{o}= e^{i \pi
/3}$,  $|b^{o}_{1}|= |b^{o}_{2}|$.  

To allow for an analytic calculation of the condensate dynamics we use a variational form of the initial state,
\begin{equation}
\tilde{\Phi}({\bf r}) = C \prod_{b^{o}}(z-b^{o})e^{-r^2/ (2d^{2})}
\label{qh} \end{equation}
where $z=x+iy$, $C$ is a normalization constant, The lengths $d$ and
$|b^{o}_{1}|$ are  variational parameters obtained by minimizing the energy $h_0+U
-\Omega L_{z}- \mu N$.  Eq.~(\ref{qh}) is the form that will emerge in the  mean
field quantum Hall regime\cite{Ho}, (in which case  $d$ is no longer be a
variational parameter but is equal to the trap length $a$ exactly). The
minimization process 
is similar to the one described in
ref.\cite{Ho}. Since our results do not depend on the
numerical value of the optimal parameters but only their  {\em general
features}, we shall only list the relevant features and not dwell on the details of the
minimization process. These features include:

\noindent {\bf (a)} A condensate with large number of vortices has rotational
frequency $\Omega\rightarrow \omega$, and  $d\rightarrow a$. 

\noindent {\bf (b)} The vortices form a triangular lattice with 
averaged vorticity $(2\pi \hbar /M)/v^{o}_{c} =
2\Omega \approx 2\omega$, which gives $v^{o}_{c} = \pi a^{2}$, and
$|b^{o}_{1}|^2=|b^{o}_{2}|^2 = 2\pi a^2/\sqrt{3}$.  

\noindent  {\bf (c)} The overall density of  $\tilde{\Phi}({\bf r})$  Gaussian
with a width $\sigma$ 
( $\sigma^2=\langle r^2 \rangle_{\tilde{\Phi}}$) such that  
$\sigma^{-2} = d^{-2} - \pi v_{c}^{-1}$. Since $d \approx a$, we have 
$\sigma \gg a$. 

\noindent {\bf (d)} 
$\tilde{\Phi}$ satisfies the lowest Landau level constraint $[(p_{x} + ip_{y})
+ \hbar z/d^2]\tilde{\Phi}=0$.   It is straightforward to show from this
condition 
that ${\cal D}$ is given by Eq.~(\ref{D}) when $\sigma^2\gg d^2$.  

 \noindent {\bf (e)}  Apart from a proportionality constant, 
$\tilde{\Phi}({\bf r})$ is the  Weierstrass $\sigma$- function\cite{Math}.

To proceed with the integration in Eq.~(\ref{int}), we use the relation between
the Weierstrass $\sigma$-function  and the $\theta$-function to write
$\tilde{\Phi}$ as \cite{MH,Math}
\begin{equation}
\tilde{\Phi}({\bf r}) =  \theta( {\bf q}_{o}\cdot {\bf r}, \tau_{o}) e^{\pi
(x+iy)^2/(2v^{o}_{c})} e^{-r^2/(2d^2)}
\label{Phitheta} \end{equation}
where ${\bf q}_{o}= (\hat{\bf x}+i\hat{\bf y})/b_{1}^{o}$, $\tau_{o} =  e^{2\pi
i/3}$, and 
\begin{equation}
\theta(\zeta, \tau) \equiv  \frac{1}{i}\sum_{n=-\infty}^{\infty} (-1)^n e^{i\pi
\tau (n+1/2)^2} 
e^{2\pi i \zeta (n+1/2)}. 
\label{thetadef} \end{equation}
We have set the  normalization constant  in Eq.~(\ref{Phitheta})  to 1 because  
it is not important  for our discussions.   To proceed further, we rewrite
$\tilde{\Phi}$ as 
\begin{eqnarray}
\tilde{\Phi}({\bf r}) &=& \theta ({\bf q}_{o} \cdot {\bf r}, \tau_{o}) e^{-
\frac{1}{2}{\bf x}\cdot {\cal M}_{o} \cdot {\bf x}}, 
\label{hexa}\\
{\cal M}_{o}&=& \left( \begin{array}[c]{cc} \frac{1}{d^{2}} -\frac{\pi}{v^{o}_{c}}
 &  -i
\frac{\pi}{v^{o}_{c}} \\
 -i \frac{\pi}{v^{o}_{c}} &   \frac{1}{d^{2}} +\frac{ \pi}{v^{o}_{c}}    \end{array} \right)
 \approx
 \frac{1}{a^2} \left( \begin{array}[c]{cc} 0 &  -i \\
-i &  2  \end{array} \right)
\label{M} \end{eqnarray}
The approximate form is due to properties ${\bf (a)}$ and ${\bf (c)}$
mentioned above.

Because of the Gaussian structure in Eqs.~(\ref{thetadef}), (\ref{hexa}), and
(\ref{green}), the integral (\ref{int}) is also a Gaussian. After sorting
out terms 
containing $(n+1/2)^2$ and $(n+1/2)$ in the exponent, $\Psi({\bf r}, t)$  can
be recast into a form similar to  Eq.~(\ref{Phitheta}), 
\begin{eqnarray}
&\Psi({\bf r},t) =  {\cal B}   \theta \left( {\bf q}_{}\cdot {\bf r}, \tau
\right) 
e^{- \frac{1}{2}{\bf x}\cdot {\cal M} \cdot {\bf x}} &
\label{scaling}\\
 &{\cal M} = {\cal P}\Lambda^{-1}{\cal P}-i{\cal Q}, \,\,\,\,\,\,\,\,
\Lambda = {\cal M}_{o} + i {\cal Q}, &
\label{calM}\\
& {\bf q} = -i {\cal P} \Lambda^{-1} {\bf q}_{o},    \,\,\,\,\,\,\,
{\bf q}_{o} =  (\hat{\bf x} + i \hat{\bf y})/b^{o}_{1},  &
\label{qqq} \\
&\tau=\tau_{o} + \eta, \,\,\,\,\,\,\,
\eta  = i 2\pi {\bf q}_{o}\cdot \Lambda^{-1} \cdot {\bf q}_{o}&
\label{eta} \\
&{\cal B} = 2\pi {\cal A}/ \sqrt{ {\rm Det} | \Lambda |}. &
\label{calB} 
\end{eqnarray}
Equations (\ref{scaling}) to ({\ref{calB}) are the key results of this paper.  
 An important feature of Eq.~(\ref{scaling}) is that {the zeros (i.e. the
vortices) of $\Psi({\bf r})$ continue to organize in a lattice during the
evolution}, with  basis vectors (${\bf b}_{1}$ and ${\bf b}_{2}$) given by 
\begin{equation}
{\bf q}\cdot {\bf b}_{1} = 1, \,\,\,\,\,\,  {\bf q}\cdot {\bf b}_{2} = \tau, 
\label{b1} \end{equation}
as a consequence of the quasiperiodicity properties, 
$\theta(\zeta+1,\tau)= - \theta(\zeta, \tau)$, and $ \theta(\zeta + \tau, \tau)
=
 - \theta(\zeta, \tau)  e^{-i\pi(\tau + 2\zeta)}$.  One can see from
Eq.~(\ref{scaling}) that the vortices follow a velocity field 
$V({\bf x})= Q_{ij} x_{j}$, where 
$Q_{ij} (t)  =\left[  (\partial_{t} {\bf b}_{1i})  {\bf K}_{1j}
+ (\partial_{t} {\bf b}_{i2} ) {\bf K}_{2j} \right] /(2\pi) $, and ${\bf
K}_{i}'s$ are the reciprocal basis vectors,  
${\bf K}_{i}\cdot {\bf b}_{j} =  2\pi \delta_{i,j}$.   
 The symmetric and antisymmetric part of $Q_{ij}$ correspond to a quadrupolar
and a rotational flow.
Thus, apart from an overall rotation, the vortices follow the streamlines of a
quadrupolar flow. 

By properties {\bf (a)} and {\bf (b)}, the parameters of
Eq.~(\ref{scaling}), are approximately
$d \approx a, |b^{o}_{1}|^2  \approx  2\pi a^2/\sqrt{3}$,
which fully specifies
${\bf q}$, $\tau$,  ${\cal M}$ and
${\cal D}$ \cite{Appendix}. While our results hold for arbitrary
$\epsilon$, we shall focus on small deformations $\epsilon<<1$ in order to
compare with current experiments\cite{JILAstripe}.  
In Fig. 2, we have plotted the density of the system at 8 equal time
intervals between $\omega t =0$ and $6$ for $\epsilon= 0.036$ and  $\Omega=
0.92\omega$.  Stripes appear periodically, becoming more distinct
at each subsequent appearance.  

To illustrate the relation between lattice deformation and stripe formation, we
find 
the images $(\overline{b}_{1}, \overline{b}_{2})$ of basis vectors  obtained
from Eq.~(\ref{b1}) in the fundamental region. The stretching of the lattice is
then describe by the ratio
 $\overline{\tau} = \overline{b}_{1}/\overline{b}_{2} = \overline{u} +
i\overline{v}$.  
Large stretching is given by large $|\overline{\tau}|$ or by large
$\overline{v}$
since $|\overline{u}|<1/2$.  
In Fig. 3, we have plotted $\overline{v}$ and the angle $\phi$ between
$\overline{b}_{1}$ and $x$ axis as a function of $\omega t$ at the lower panel. 
We find that (i) stripes are most pronounced when
$\overline{v}$ is a maximum, which occurs six times within a period $\omega t$.
Moreover, the sharpness of stripe density increases with time; 
(ii)  the maximum occurs when $\phi$ is $45^{o}$, which is the direction of the
minor axis after half of the period as shown in Fig. 1b.  These two detailed
features are in exact agreement with those observed in ref.\cite{JILAstripe}. 
The discontinuities in $\phi$ are due to the mapping of the basis vectors back
into the fundamental region when the lattice is stretched.  The orientation of
the second basis vector $\overline{b}_{2}$ is given in the top panel of fig.3, where
we plot the relative angle $\kappa$ between $\overline{b}_{1}$ and 
$\overline{b}_{2}$.  Once again, the discontinuities of this angle is due to the
mapping back to the fundamental region.  
To illustrate the {\em continuous} evolution of the lattice  and the 
discontinuous change of the fundamental basis, we have plotted the lattices and
indicated their fundamental basis at different times in Fig. 4. 
It is also interesting to note that at the cusps in fig. 3, the unit cell of
vortex lattice is close to a square.

{\bf (III) The origin of stripe formation:}  Figure 3 and 4 show that vortex
stripes can be produced by rather moderate changes in the basis vectors. 
This sensitivity is
due to the special coherent properties of the vortex lattice, as revealed
by studying the periodic part of the density which is calculated from
\cite{MH}, 
\begin{equation}
|\theta(\zeta, \tau)|^2= \left[ \frac{1}{v_{c}} \sum_{\bf K} g_{\bf K}e^{i{\bf
K}\cdot {\bf r}} 
\right] e^{2\pi y^2/v_{c}} 
\end{equation}
where ${\bf K}=$$m_{1}{\bf K}_{1}$$+m_{2}{\bf K}_{2}$ are the reciprocal
lattice vectors, $m_{i}$ are integers, and  $ g_{\bf K}$ $ = (-1)^{m_{1} +
m_{2} + m_{1}m_{2}}$ $e^{-v_{c} |{\bf K}|^2/8\pi}$$\sqrt{v_{c}/2}$. Because of
the rapid decay of $g_{\bf K}$ with increasing $|{\bf K}|$,  only a few
${\bf K}$ vectors around the origin have non-trivial contribution to the
density.   For a triangular lattice, it is natural to organize the ${\bf K}-points$ into concentric hexagonal rings, corresponding to reciprocal lattice vectors of roughly equal length, and to number these rings, starting from the origin. 
The same labeling works for distorted lattices, and for sufficiently small distortions
the first ring, made up of the six shortest wavevectors, has the largest contribution to the density.  
As $|\tau|= |b_{2}/b_{1}|$ increases, $b_{1}$ becomes the shortest vector in
real space, which in turn makes $\pm {\bf K}_{1}$ the shortest vector in
k-space, and this wave-vector dominates the contribution from the first ring.  Similarly,
${\pm}2{\bf K}_{1}$ dominates the contribution from the second ring, and so on.
As a result,  the density is increasingly
dominated by the sum of $g_{\bf K}$'s  containing only integer multiples of
${\bf K}_{1}$, which gives rise to  a sequence of stripes in real space.  

{\bf Final Remarks:}  We have shown that the mysterious periodic stripe 
formation in ref.\cite{JILAstripe} reflects the kinetic evolution of the vortex
lattice, where vortices remained organized in lattice,  with each vortex
following the streamline of a quadrupolar flow. The six occurrences of stripes
during a period is caused by an almost periodic deformation of the vortex
lattice and not the melting of vortices into sheets. 
The method presented here should also be applicable to other to dynamical
phenomena at high angular momenta and to more complex condensates
\cite{Gardina}.

This work is supported by NASA Grants NAG8-1441, NAG8-1765, and by NSF Grants
DMR-0109255, DMR-0071630.

\begin{figure}[b]
\psfrag{m}[bl]{$\pi$}
\psfrag{la}[tc]{$\lambda_-^{3/2} \lambda_+^{1/2}$}
\psfrag{as}[cl]{\quad$\displaystyle\sqrt{\frac{\lambda_-}{\lambda_+}}$}
\psfrag{ch}[bl]{$\chi$}
\psfrag{mf}[br]{$\pi/4$}
\psfrag{wt}[bl]{$\omega t$}
\includegraphics[width=\columnwidth]{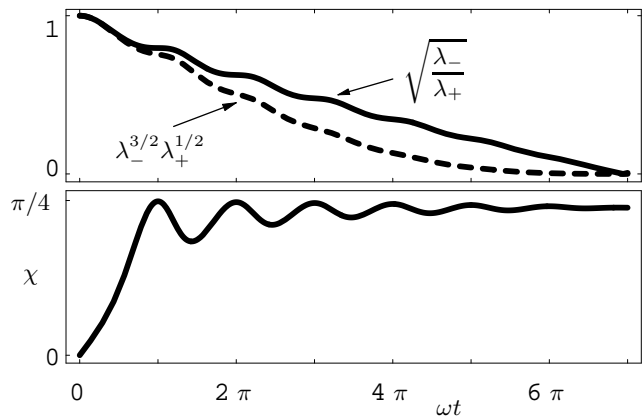}
\caption{(a) The product $\lambda^{3/2}_{-}\lambda^{1/2}_{+}$ and 
the aspect ratio $\sqrt{\lambda_{-}\lambda_{+} }$ for $\epsilon = 0.036$.
(b) The angle $\chi$ between the minor axis $\hat{\bf n}_{-}$ and the
$x$-axis.}
\end{figure}

\begin{figure}[b]
\includegraphics[width=\columnwidth]{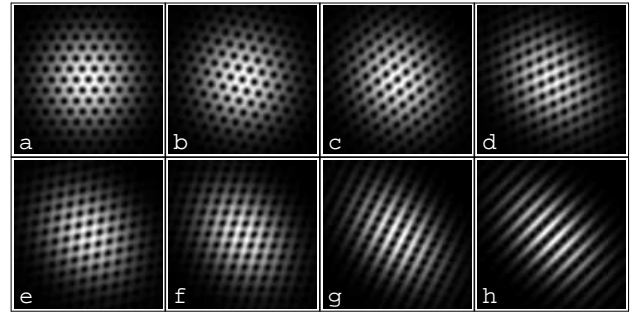}
\caption{Density profile of the evolving condensate from $\omega
t=0$ to $6$ at equal time interval. Dark color means low density. 
Panel (h) corresponds to point (c) on Fig.~\ref{geo}.
}
\end{figure}

\begin{figure}[b]
\psfrag{L}{$\bar v$}
\psfrag{w}{$\omega t$}
\psfrag{t}{}
\psfrag{x}{$\kappa$}
\psfrag{f}{$\phi$}
\includegraphics[width=\columnwidth]{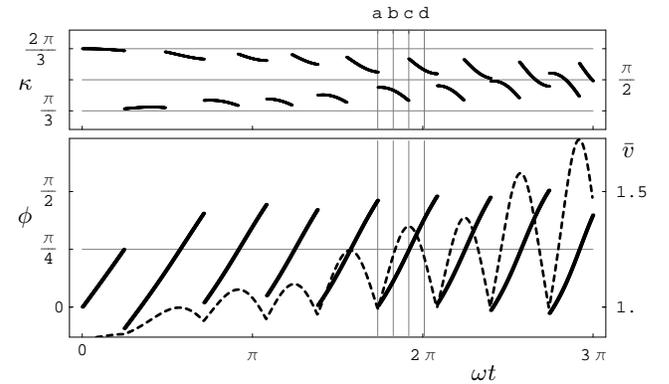}
\caption{The lower panel shows $\overline{v}$ (dotted line) and the
angle 
$\phi$ between $\overline{b}_{1}$ and the $x$-axis as a function of $\omega t$.
The peaks of $\overline{v}$ correspond to the appearance of stripes, and
coincide with $\phi=\pi/4$; ie. $\bf b_1$ aligned with the minor axis.
The upper panel shows the angle $\kappa$ between the vectors $\overline{b}_{2}$
and $\overline{b}_{1}$. The lattice configuration and basis vectors at times 
$(a), (b), (c), (d)$  are shown in Fig. 4. }\label{geo}
\end{figure}

\begin{figure}[b]
\psfrag{H}{}
\psfrag{L}{}
\psfrag{a}[bc]{(a)}
\psfrag{b}[bc]{(b)}
\psfrag{c}[bc]{(c)}
\psfrag{d}[bc]{(d)}
\includegraphics[width=\columnwidth]{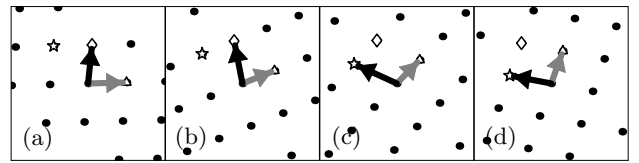}
\caption{The lattice configurations and the basis vectors in the
fundamental region at different time $(a), (b), (c), (d)$ indicated in Fig.
3. The discrete change of basis vectors between $(b)$ and $(c)$ is due to the
mapping back to the fundamental region. }
\end{figure}
\clearpage



\begin{thebibliography}{99}
\bibitem{comment1} The large response is due to the fact that one is near the
resonance of the $m=-2$ mode.
\bibitem{Appendix} As an illustration, for times $2\epsilon \omega t<<1$, 
which include $\omega t>1$, we have 
$\eta = \epsilon \sqrt{3} e^{-2i\omega t}\left(
-2\omega t + {\rm sin}2\omega t \right) +O(\epsilon^2)$ from Eq.~(\ref{eta}), 
and  ${\bf q} =  \nu(t) (\hat{\bf x}+ i\hat{\bf y})/b^{o}_{1} + \xi(t)
(\hat{\bf x}- i\hat{\bf y}) /b^{o}_{1} $ 
from Eq.~(\ref{qqq}), with 
$\xi=- i \epsilon e^{-2 i \omega t}\left[
         e^{i \omega t} \omega t - \sin(\omega t)\right]$,
and
$
\nu=e^{-3 i \omega t} \left[
e^{2 i \omega t} + i\epsilon (2 \omega t - \sin(2\omega t))
\right]$.
If we define $q_{\pm} = {\bf q}\cdot (\hat{\bf x}\pm i \hat{\bf y})$, then
$q_{+}= 2\xi/b_1^o$, $q_{-}=2\nu/b_1^o$, and Eq.~(\ref{b1}) implies
 $b_{1}q_{-}+ b_{1}^{\ast} q_{+} = 2$. We then have 
$b_{1}/b_1^o =
e^{i\omega t}-i \epsilon [e^{-i \omega t} \omega t- \sin(\omega t)] +O(\epsilon^2)$.
 All our calculations,
however, are performed with the exact expressions Eq.~(\ref{calM}) to
(\ref{calB}) without expansions.
\bibitem{Gardina} Professor Crispin Gardiner recently informed us that he has
also found stripes in a numerical simulation of a modified GP equation which
includes dissipation.
\end{thebibliography}
\end{document}